\documentclass[aps,pra,twocolumn,showpacs,preprintnumbers,amsmath,amssymb,superscriptaddress,nofootinbib,balancelastpage,showkeys]{revtex4}

\usepackage{amsmath,amssymb}
\usepackage[latin1]{inputenc}
\usepackage{dcolumn}
\usepackage{bm}

\usepackage{graphicx}% Include figure files
\usepackage{dcolumn}% Align table columns on decimal point
\usepackage{bm}% bold math
\begin{document}

%\preprint{APS/123-QED}

\title{Faked states attack using detector efficiency mismatch on SARG04, phase-time, DPSK, and Ekert protocols}% Force line breaks with \\

\author{Vadim Makarov}
\email{makarov@vad1.com}
\affiliation{Department of Electronics and Telecommunications, Norwegian University of Science and Technology, NO-7491 Trondheim, Norway}
\affiliation{Radiophysics Department, St.~Petersburg State Polytechnic University, Politechnicheskaya street 29, 195251 St.~Petersburg, Russia}
\affiliation{Department of Physics, Pohang University of Science and Technology, Pohang 790-784, South Korea}

\author{Johannes Skaar}
\affiliation{Department of Electronics and Telecommunications, Norwegian University of Science and Technology, NO-7491 Trondheim, Norway}

\date{November~23, 2007}% It is always \today, today,

\begin{abstract}
In quantum cryptosystems, variations in detector efficiency can be exploited to stage a successful attack. This happens when the efficiencies of Bob's two detectors are different functions of a control parameter accessible to Eve (e.g., timing of the incoming pulses). It has previously been shown that the Bennett-Brassard 1984 (BB84) protocol is vulnerable to this attack. In this paper, we show that several other protocols and encodings may also be vulnerable. We consider a faked states attack in the case of a partial efficiency mismatch on the Scarani-Acin-Ribordy-Gisin 2004 (SARG04) protocol, and derive the quantum bit error rate as a function of detector efficiencies. Additionally, it is shown how faked states can in principle be constructed for quantum cryptosystems that use a phase-time encoding, the differential phase shift keying (DPSK) and the Ekert protocols.
\end{abstract}

% PACS, the Physics and Astronomy Classification Scheme.
%Use showkeys class option if keyword display desired
\maketitle

\section{\label{sec:intro}Introduction}

Quantum key distribution (QKD) is a technique that allows remote parties to grow shared secret random key at a steady rate, given an insecure optical communication channel and an initially authenticated classical communication channel between them \cite{bennett1992,quantum-cryptography-reviews}. Since the first experimental demonstration eighteen years ago \cite{bennett1992}, QKD systems have developed to commercial devices working over tens of kilometers of optical fiber \cite{commercial-QKD-systems}, as well as experiments over more than a hundred kilometers of fiber \cite{NaturePhotonics-1-p343,NewJPhys-8-p193,JpnJApplPhys-43-pL1217,ApplPhysLett-84-p3762,takesue_2005,over-100km-decoy}, 23~km and 144~km of free space \cite{PhysRevLett-98-p010504,NaturePhysics-adv-nphys629,Nature-419-p450-and-ProcSPIE-4917-p25}. Although the security of QKD has been unconditionally proven for a model of equipment that includes certain non-idealities \cite{gott_lo_lutk_presk,shor_preskill,koashi_preskill,Proc32ndAnnualACMSympTheoryComp-p715,EurPhysJD-41-p599}, not all real properties of optical and electrooptical components have been included into the proof. Identifying the properties of components potentially dangerous for security and integrating them into the proof (or closing the issue in some other way) is an ongoing work \cite{JModOpt-48-p2039,large-pulse-attack,makarov,PhysRevA-74-p022313,QuantInfComp-7-p73,arXiv-0704-3253v1-quant-ph,OptExpress-15-p9388}.

In this paper, we continue to analyse a common imperfection of Bob's single photon detectors: variation of their efficiency that can be controlled by Eve via a choice of an external parameter. It has been shown in Refs.~\onlinecite{PhysRevA-74-p022313} and \onlinecite{QuantInfComp-7-p73} that even smallest variations of one detector efficiency relative to the other detector reduce the amount of secret information theoretically available to Alice and Bob in the case of the BB84 protocol. The amount of key compression during the privacy amplification must be adjusted based on an evaluation of the worst-case efficiency mismatch of Bob's detectors. We recap these results in Sec.~\ref{sec:BB84}. In the following sections, we consider other protocols and encodings: SARG04 in Sec.~\ref{sec:SARG04}, a class of schemes using the phase-time encoding and the DPSK protocol in Sec.~\ref{sec:phase-time}, and the Ekert protocol with a source of entangled photons in Sec.~\ref{sec:Ekert}. It is shown how to construct a faked states attack \cite{makarov} against these protocols and encodings. For the SARG04, the upper bound on available secret key information is estimated, through calculating the quantum bit error rate (QBER) caused by this attack in the case of a partial efficiency mismatch. For the other protocols, we consider the case of a total efficiency mismatch only and make no quantitative estimates. Although the case of the total efficiency mismatch can occur in practice \cite{PhysRevA-74-p022313}, usually detectors in a QKD system will merely have some partial efficiency mismatch. This work is thus the first step in analysing detector efficiency mismatch in these protocols.

\section{\label{sec:BB84}BB84 protocol}

Variation of efficiency is a common and, indeed, unavoidable imperfection of single photon detectors. The efficiency may depend on the timing of incoming light pulse (e.g., in gated detectors based on avalanche photodiodes), wavelength of incoming light (e.g., in up-conversion detectors \cite{thew_2006,langrock_2005,diamanti_2005}), polarization and other parameters conceivably controllable by Eve. In QKD schemes that employ two detectors (or a time-multiplexed detector), the variation will be different between the detectors (or detection windows), allowing Eve to control the relative probability of one detection outcome over the other.

To illustrate how she can use this to construct a successful attack on the BB84 protocol \cite{bennett1992}, we assume at first that the efficiency mismatch for some values of the control parameter is so large that Eve can practically blind either detector while the other remains sensitive. This situation is called a {\em total efficiency mismatch}. We call the value of the control parameter that blinds the 1 detector $t_0$, and the value that blinds the 0 detector $t_1$.

Eve then proceeds with an intercept-resend attack: she uses a replica of Bob's setup to detect every Alice's state, and resends certain states of light to Bob. It is well known that a straightforward intercept-resend attack, in which Eve resends quantum states that simply repeat her detection results (bit value and basis), is doomed to fail. This is because Eve does not know Alice's basis, will thus detect half of Alice's qubits in a wrong basis, and cause 25\% errors in Bob's key. However, our intercept-resend attack has an important twist: Eve sends states of light that only get detected by Bob when he chooses {\em the same basis as Eve,} otherwise they cause no click in Bob's detectors (we'll explain in a moment how Eve achieves this). In such a case, all Eve's detections in a wrong basis belong to the qubits detected by Bob in the same wrong basis, and are discarded by Alice and Bob during sifting. What remain after sifting are those bits which have been sent by Alice, detected by Eve and detected by Bob in the same basis for all three parties. This key is error-free, and Eve knows every bit of it.

The intercept-resend attack ``with a twist'' described above is a faked states attack, and the specially formed light states Eve resends to Bob are called faked states \cite{makarov}. The faked state Eve resends in our case would be a state normally used in the protocol but with the opposite bit value in the opposite basis comparing to what she has detected. At the same time, in the faked state Eve sets the value of the control parameter that blinds the detector for the opposite bit value from what she has detected. For example, suppose Eve has detected the 0 bit value in the X basis. She resends the 1 bit in the Z basis, with the control parameter $t_0$. If Bob tries to detect this faked state in the Z basis, he never detects anything, for his 1 detector is blinded by Eve's choice of the control parameter. If he tries to detect in the X basis, he with equal probability doesn't detect anything or detects the 0 bit. The reader may notice that the attack reduces the detection probability at Bob, but this can be compensated by a proportionally increased brightness of the faked states. Thus, in the case of the total efficiency mismatch, Eve can run a faked states attack that causes zero QBER and gives her full information on the key \cite{PhysRevA-74-p022313}.

In the case of a {\em partial efficiency mismatch}, when either detector cannot be completely blinded, this attack causes some non-zero QBER. Eve can pick the values of the control parameter to minimize the ratios $\eta_0(t_1)/\eta_1(t_1)$ and $\eta_1(t_0)/\eta_0(t_0)$, where $\eta_0$ and $\eta_1$ are efficiencies of the 0 and 1 detectors. It has been shown in Ref.~\onlinecite{PhysRevA-74-p022313} that in this case the attack causes 
\begin{equation}
\label{eq:BB84_QBER}
\text{(QBER)}=\frac{2\eta_0(t_1)+2\eta_1(t_0)}{\eta_0(t_0)+3\eta_0(t_1)+3\eta_1(t_0)+\eta_1(t_1)}.
\end{equation}
In the special case of symmetric detector efficiency curves $\eta_0(t_1)/\eta_1(t_1) = \eta_1(t_0)/\eta_0(t_0) \equiv \eta$
and Eve adjusting the brightness of her faked states sent with $t_0$ and $t_1$ such that Bob's detection probability for both values of the control parameter remains equal, this simplifies to
\begin{equation}
\label{eq:BB84_QBER_symm}
\text{(QBER)}=\frac{2\eta}{1+3\eta}.
\end{equation}
The QBER value of 0.11 (commonly regarded as the threshold value for the BB84 protocol, after which no secret key could be extracted) would be reached at $\eta \approx 1/15$.

The attack described above is not necessarily optimal. In Ref.~\onlinecite{PhysRevA-74-p022313} we say that the BB84 protocol is secure provided $(\text{QBER}) \alt 0.11 \eta$ and an extra amount of privacy amplification is applied. However, it has since been noticed that Eq.~11 in Ref.~\onlinecite{PhysRevA-74-p022313}, on which this conclusion is based, is incorrect. It follows from Eq.~11 that if QBER is zero, Eve has no information. Qi and coworkers have pointed out that when Eve can affect Bob's detector efficiencies, she gets partial information about the key from Bob's announcement of which qubits have actually arrived \cite{QuantInfComp-7-p73}. Thus the available bit rate after privacy amplification is reduced even in the case $(\text{QBER})=0$. This makes possible the so-called time-shift attack, in which Eve alters randomly the control parameter of the qubits without otherwise interacting with them \cite{QuantInfComp-7-p73,arXiv-0704-3253v1-quant-ph}. A purely classical side-channel attack on a system where Bob measures and announces detection timing has also been proposed \cite{OptExpress-15-p9388}. A more general theory, which is not yet available, would encompass the time-shift (or, more generally, parameter-shift) attacks into the equation for the available bit rate.

\section{\label{sec:SARG04}SARG04 protocol}

\begin{figure}
\includegraphics{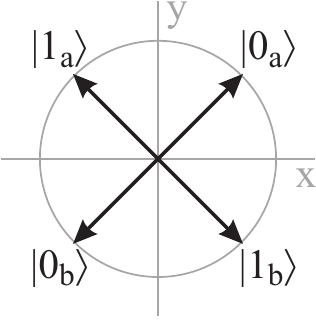}
\caption{\label{fig:sarg04-states}States configuration for the SARG04 protocol in the case when the states used are physically equivalent to those in the BB84 protocol. The circle represents the equator of the Poincare sphere.}
\end{figure}

The purpose of the SARG04 protocol \cite{scarani_acin,acin_gisin,branciard_2005} is to increase the maximum trasmission distance and key yield in schemes that use a weak coherent source; the protocol has improved characteristics against the photon number splitting attack, comparing to the BB84. Here we consider the version of the SARG04 that uses states physically equivalent to those used in the BB84 (Fig.~\ref{fig:sarg04-states}),
and differs from the latter only at the sifting stage. The bit values 0 and 1 in the SARG04 are encoded by the choice of basis. Alice sends randomly one of the four states $|0_a\rangle$, $|0_b\rangle$, $|1_a\rangle$ or $|1_b\rangle$. Bob measures either in the 0 or 1 detection basis, and uses two detectors labeled $a$ and $b$. At the sifting stage, Alice announces publicly a set of two states that contains the actual state sent and a random state from the opposite basis. For definiteness, suppose that Alice has sent $|0_a\rangle$ and that she has announced the set $\{|0_a\rangle,|1_a\rangle\}$. If Bob has measured in the 0 basis, he has certainly got the result $0_a$; but since this result is possible for both states in the set $\{|0_a\rangle,|1_a\rangle\}$, he has to discard it. If he has measured in the 1 basis and got $1_a$, he again cannot discriminate. But if he has measured in the 1 basis and got $1_b$, he knows that Alice has sent $|0_a\rangle$, and adds 0 to his key.

Since this protocol uses the same states as the BB84, the faked states attack described in the previous section could be applied to it. In the case of the total efficiency mismatch, it obviously causes zero QBER. To calculate the QBER it causes in the case of the partial efficiency mismatch, we follow the approach of Ref.~\onlinecite{PhysRevA-74-p022313} and consider all the possible basis and detector combinations during the attack. The different events are shown in Table~\ref{table:SARG04_attack}
\begin{table}[t]
\newcommand\Ta{\rule{0pt}{2.5ex}}
\newcommand\Tb{\rule{0pt}{3.0ex}}
\caption{\label{table:SARG04_attack}The intercept-resend attack on the SARG04 protocol when Alice sends the $|0_a\rangle$ state (as indicated in the first column; in the table, brackets around states are omitted for clarity). The second column contains the basis chosen by Eve and the measurement result; the third column shows the state and timing as resent by Eve. In the next columns Bob's basis choice and measurement results are given. For the case with the partial detector sensitivity mismatch, the probabilities for the different results are shown, given Eve's state and timing in addition to Bob's basis. In the last two columns, pairs of states announced by Alice during sifting (two possible pairs announced with equal probability of 1/2), and the sifting results, are shown. Note that, for ease of discussion, the first two rows are repeated so that each row in the table occurs with probability 1/8.\smallskip}
\begin{ruledtabular}
\begin{tabular}{@{}c c c c c c c c@{}}
Alice & $\rightarrow$Eve & Eve$\rightarrow$ & Bob & \multicolumn{2}{l}{\begin{tabular}{@{}l}\Ta Result,\\ \ \ \ \  Probability\\[0.8ex]\end{tabular}} & \begin{tabular}{@{}c@{}}\Ta Alice's\\announce\\[0.8ex]\end{tabular} & Sifting\\[0.7ex]
\hline
\Tb
$0_a$ & $0_a$ & $1_b t_a$ & $0$ & $a$, & $\frac{1}{2}\eta_a(t_a)$ & $\{0_a,1_a\}$ & Discard      \\[0.4ex]
      &       &           &     &      &                          & $\{0_a,1_b\}$ & Discard      \\[0.9ex]
      &       &           &     & $b$, & $\frac{1}{2}\eta_b(t_a)$ & $\{0_a,1_a\}$ & $1_a$ (error)\\[0.4ex]
      &       &           &     &      &                          & $\{0_a,1_b\}$ & $1_b$ (error)\\[1.4ex]

$0_a$ & $0_a$ & $1_b t_a$ & $1$ & $a$, & $0$                                                     \\[0.9ex]
      &       &           &     & $b$, &        \ \ $\eta_b(t_a)$ & $\{0_a,1_a\}$ & $0_a$ (right)\\[0.4ex]
      &       &           &     &      &                          & $\{0_a,1_b\}$ & Discard      \\[1.4ex]

$0_a$ & $0_a$ & $1_b t_a$ & $0$ & $a$, & $\frac{1}{2}\eta_a(t_a)$ & $\{0_a,1_a\}$ & Discard      \\[0.4ex]
      &       &           &     &      &                          & $\{0_a,1_b\}$ & Discard      \\[0.9ex]
      &       &           &     & $b$, & $\frac{1}{2}\eta_b(t_a)$ & $\{0_a,1_a\}$ & $1_a$ (error)\\[0.4ex]
      &       &           &     &      &                          & $\{0_a,1_b\}$ & $1_b$ (error)\\[1.4ex]

$0_a$ & $0_a$ & $1_b t_a$ & $1$ & $a$, & $0$                                                     \\[0.9ex]
      &       &           &     & $b$, &        \ \ $\eta_b(t_a)$ & $\{0_a,1_a\}$ & $0_a$ (right)\\[0.4ex]
      &       &           &     &      &                          & $\{0_a,1_b\}$ & Discard      \\[1.4ex]

$0_a$ & $1_a$ & $0_b t_a$ & $0$ & $a$, & $0$                                                     \\[0.9ex]
      &       &           &     & $b$, &        \ \ $\eta_b(t_a)$ & $\{0_a,1_a\}$ & $1_a$ (error)\\[0.4ex]
      &       &           &     &      &                          & $\{0_a,1_b\}$ & $1_b$ (error)\\[1.4ex]

$0_a$ & $1_a$ & $0_b t_a$ & $1$ & $a$, & $\frac{1}{2}\eta_a(t_a)$ & $\{0_a,1_a\}$ & Discard      \\[0.4ex]
      &       &           &     &      &                          & $\{0_a,1_b\}$ & $0_a$ (right)\\[0.9ex]
      &       &           &     & $b$, & $\frac{1}{2}\eta_b(t_a)$ & $\{0_a,1_a\}$ & $0_a$ (right)\\[0.4ex]
      &       &           &     &      &                          & $\{0_a,1_b\}$ & Discard      \\[1.4ex]

$0_a$ & $1_b$ & $0_a t_b$ & $0$ & $a$, &        \ \ $\eta_a(t_b)$ & $\{0_a,1_a\}$ & Discard      \\[0.4ex]
      &       &           &     &      &                          & $\{0_a,1_b\}$ & Discard      \\[0.9ex]
      &       &           &     & $b$, & $0$                                                     \\[1.4ex]

$0_a$ & $1_b$ & $0_a t_b$ & $1$ & $a$, & $\frac{1}{2}\eta_a(t_b)$ & $\{0_a,1_a\}$ & Discard      \\[0.4ex]
      &       &           &     &      &                          & $\{0_a,1_b\}$ & $0_a$ (right)\\[0.9ex]
      &       &           &     & $b$, & $\frac{1}{2}\eta_b(t_b)$ & $\{0_a,1_a\}$ & $0_a$ (right)\\[0.4ex]
      &       &           &     &      &                          & $\{0_a,1_b\}$ & Discard      \\[1.6ex]
\end{tabular}
\end{ruledtabular}
\end{table}
for the special case where Alice sends the $|0_a\rangle$ state (the other three cases are symmetrical to this case). We disregard the probability of Eve's and Bob's detectors firing simultaneously due to the multiphoton fraction of the pulses, assume that Bob's detectors have no dark counts, assume that Eve's detectors and optical alignment are perfect, and that Eve generates faked states that match the optical alignment in Bob's setup perfectly. None of these assumptions is critical for the attack to work, but it is convenient to make them in order to simplify the calculation.

Based on the probabilities in the table, we calculate the QBER caused by the attack. When Alice sends the $|0_a\rangle$ state, the probability that the qubit arrives at Bob and is {\em not} discarded as an inconclusive detection result during sifting is
\begin{equation}
\label{SARG04_Parr0a}
P(\text{arrive}|A\text{=}0_a\!)\text{=}\frac{1}{8}[\frac{1}{4}\eta_a(t_a)+\frac{1}{4}\eta_a(t_b)+\frac{13}{4}\eta_b(t_a)+\frac{1}{4}\eta_b(t_b)].
\end{equation}
%while the error probability is
%\begin{equation}
%\label{SARG04_Perr0a}
%P(\text{error}|\text{A=}0_a)=\frac{1}{8}[2\eta_b(t_a)].
%\end{equation}
The probability of arrival averaged over Alice's four state choices is found by symmetrization of this equation, yielding
\begin{equation}
\label{SARG04_Parr}
P(\text{arrive})=\frac{1}{32}[\eta_a(t_a)+7\eta_a(t_b)+7\eta_b(t_a)+\eta_b(t_b)].
%\label{SARG04_Perr}
%P(\text{error})\ =\frac{1}{32}[4\eta_b(t_a)+4\eta_a(t_b)].
\end{equation}
Similarly, we find the QBER,
\begin{equation}
\label{eq:SARG04_QBER}
\text{(QBER)}\text{=}\frac{P(\text{error})}{P(\text{arrive})}\text{=}\frac{4\eta_a(t_b)+4\eta_b(t_a)}{\eta_a(t_a)+7\eta_a(t_b)+7\eta_b(t_a)+\eta_b(t_b)},
\end{equation}
where $P(\text{error})$ accounts for the cases when Bob keeps a bit value different from what Alice has sent.

In the special case of symmetric detector efficiency curves,
%$\eta_a(t_b)/\eta_b(t_b) = \eta_b(t_a)/\eta_a(t_a) \equiv \eta$ and Eve adjusting brightness of her faked states sent with timing $t_a$ and $t_b$ such that Bob's detection probability for both timings stays the same
we get
\begin{equation}
\label{eq:SARG04_QBER_symm}
\text{(QBER)}=\frac{4\eta}{1+7\eta}.
\end{equation}

Security bounds and operating conditions for the SARG04 and BB84 protocols are different \cite{branciard_2005,PhysRevA-73-p010302R,quant-ph-0507154,PhysRevA-73-p012337,PhysRevA-75-p032341}. The same value of optical misalignment (measured by, e.g., fringe visibility in the interferometer) leads to different QBER for the two protocols. The optimal photon number in a weak-pulse implementation differs between the protocols, so detector dark counts will make a different contribution to the QBER as well \cite{branciard_2005,PhysRevA-73-p012337}. Therefore, a system using the same optical hardware and the same communication line will run at a different QBER level for each protocol. If we wanted to compare the QBER caused by our attack on these two protocols, it should be done in this context, which is not at all straightforward. We note, however, that in SARG04 our attack causes QBER lower than 0.11 when $\eta \alt 1/30$, while in BB84 (see Eq.~\ref{eq:BB84_QBER_symm}) the same happens when $\eta \alt 1/15$. The effect of the described attack on these two protocols appears to be of the same order of magnitude.

The faked states attack leads to reduced bit rate according to Eq.~\ref{SARG04_Parr}. Eve may compensate this by resending a brighter signal. Alternatively, she may place her measurement device close to Alice and her resend device close to Bob, getting rid of the channel loss. The attack may also lead to altered coincidence count rates at Bob. However, with the help of timing and state parameters, Eve may have several degrees of freedom to compensate this as well. For example, by eliminating the channel loss and resending single photons, the coincidence detections may be eliminated. Furthermore, the ability to do photon number measurement on Alice's pulses would allow Eve to completely remove her influence on coincidence counts \cite{PhysRevA-74-p022313}. She could, for instance, only attack single-photon pulses, while passing multi-photon ones (those that cause coincidence counts) to Bob undisturbed, at the cost of not getting a small fraction of the key. How much Eve would have to do in practice depends, of course, on the actual checks Bob implements (or not implements, as may be the case).

\section{\label{sec:phase-time}Phase-time encoding and DPSK protocol}

\begin{figure*}
\includegraphics[width=125mm]{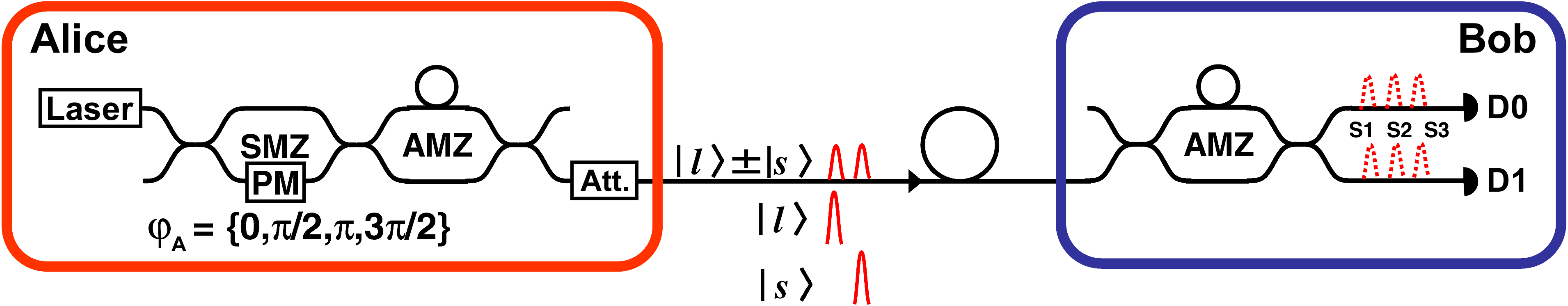}
\caption{\label{fig:phase-time-scheme}Scheme of a QKD system utilizing the phase-time encoding \cite{nambu_2004}. SMZ, symmetric Mach-Zehnder interferometer; AMZ, asymmetric Mach-Zehnder interferometer; PM, phase modulator; Att., optical attenuator; D0 and D1, single photon detectors.}
\end{figure*}

In a QKD system with the phase-time encoding \cite{nambu_2004}, Alice prepares one of the four states: $|l\rangle$, $|s\rangle$,  $|l\rangle+|s\rangle$ or $|l\rangle-|s\rangle$, where $|l\rangle$ and $|s\rangle$ denote states that have travelled via the long and short arm of Alice's AMZ (Fig.~\ref{fig:phase-time-scheme}).
Bob gates his detectors three times. The state $|l\rangle$ can cause a detection either in the S1 or S2 time slot. The state $|s\rangle$ can cause a detection either in the S2 or S3 time slot. The states $|l\rangle+|s\rangle$ and $|l\rangle-|s\rangle\}$ can cause a detection in any of the three time slots. The plus or minus sign determines which of the two detectors (D0 or D1) clicks when the detection happens in the S2 time slot where the pulses from the two arms of Bob's AMZ have interfered. Thus, pairs of states $\{|l\rangle,|s\rangle\}$ and $\{|l\rangle+|s\rangle,|l\rangle-|s\rangle\}$ form two bases. This system uses the BB84 protocol. (We note that the function of Bob's apparatus is similar to an earlier system that uses entangled photons in energy-time Bell states \cite{tittel_2000}.)

Faked states for this QKD system are listed in Table~\ref{table:phase-time-faked-states}.
Eve uses an apparatus that can form a single pulse (denoted $|ll\rangle$) in the time slot that follows the time slot of Alice's $|l\rangle$ state, a single pulse (denoted $|ss\rangle$) in the time slot that precedes the time slot of Alice's $|s\rangle$ state, or coherent states consisting of four pulses with certain phase shifts between them and a certain value of the control parameter $t$ (which can be timing as shown on the diagrams, or some other parameter). The single pulse states are sent with the control parameter value $t_\text{normal}$ that blinds neither detector. The coherent four pulse states are sent with the control parameter value $t_0$ or $t_1$ that blinds the detector D1 or D0. The faked states rely on the lack of detector gating in what would be Bob's time slots S0 and S4, or on Bob discarding detection results with these times. Additionally, in the last two faked states, Eve blinds one of Bob's detectors by the choice of the control parameter.

\begin{table}[t]
\caption{\label{table:phase-time-faked-states}Faked states for a QKD system utilizing the phase-time encoding. Each faked state is illustrated by a time diagram. The arrows indicate how every pulse coming to Bob is split into the two arms of his interferometer. The waveform for the intensity of light at Bob's detector that is blinded by Eve's choice of the control parameter $t$ is printed in gray.\medskip}
\begin{ruledtabular}
\begin{tabular}{@{}c@{}}
\includegraphics[width=87mm]{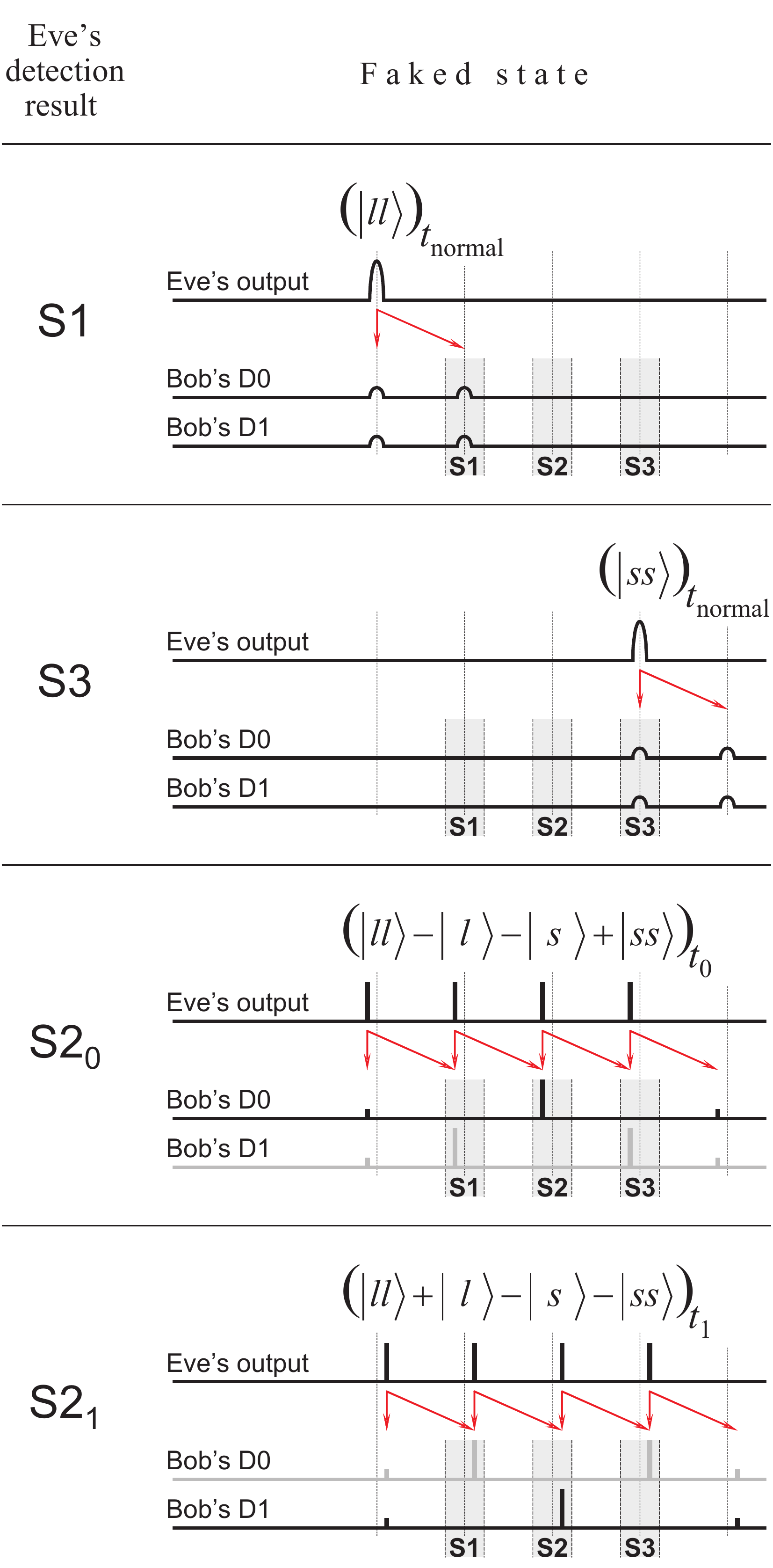}\\[1.08ex]
\end{tabular}
\end{ruledtabular}
\end{table}

\begin{figure*}
\includegraphics[width=100mm]{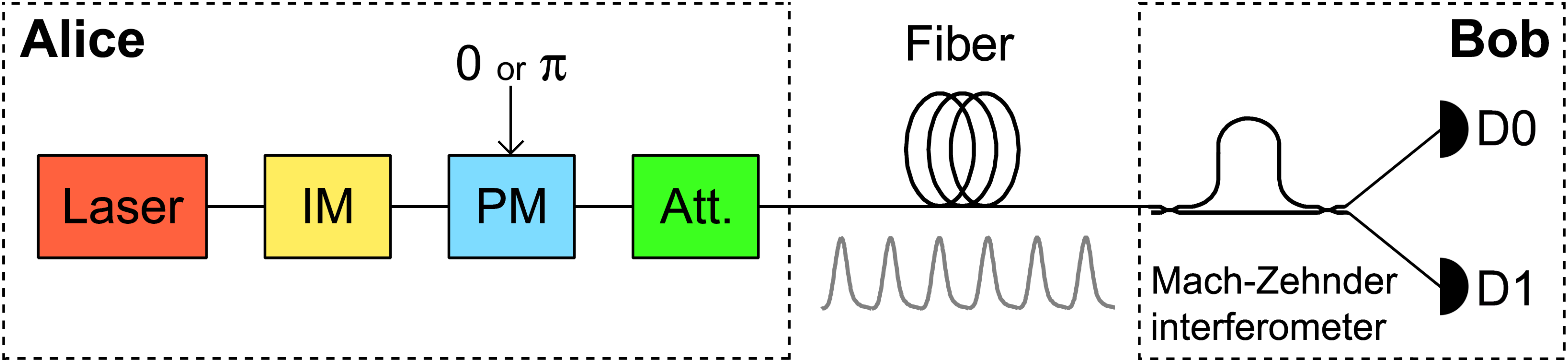}
\caption{\label{fig:DPSK-scheme}Scheme of a QKD system utilizing the DPSK protocol \cite{takesue_2005}. IM, intensity modulator; PM, phase modulator; Att., optical attenuator; D0 and D1, single photon detectors.}
\end{figure*}

In a QKD system with the DPSK protocol \cite{takesue_2005}, Alice randomly modulates the phase of a weak coherent pulse train by $\{0,\pi\}$ for each pulse, and sends it to Bob with an average photon number of less than 1 per pulse (Fig.~\ref{fig:DPSK-scheme}).
Bob measures the phase difference between adjacent pulses with a \hbox{1-bit} delay interferometer followed by two detectors placed at the interferometer output ports. Detector D0 clicks when the phase difference is 0 and detector D1 clicks when the phase difference is $\pi$. Since the average photon number per pulse is less than 1, Bob observes clicks only occasionally and in a random time slot. Bob informs Alice of the time slots in which he has observed clicks. From her modulation data Alice knows which detector has clicked on Bob's side, so they share an identical bit string.

\begin{figure*}
\includegraphics[width=150mm]{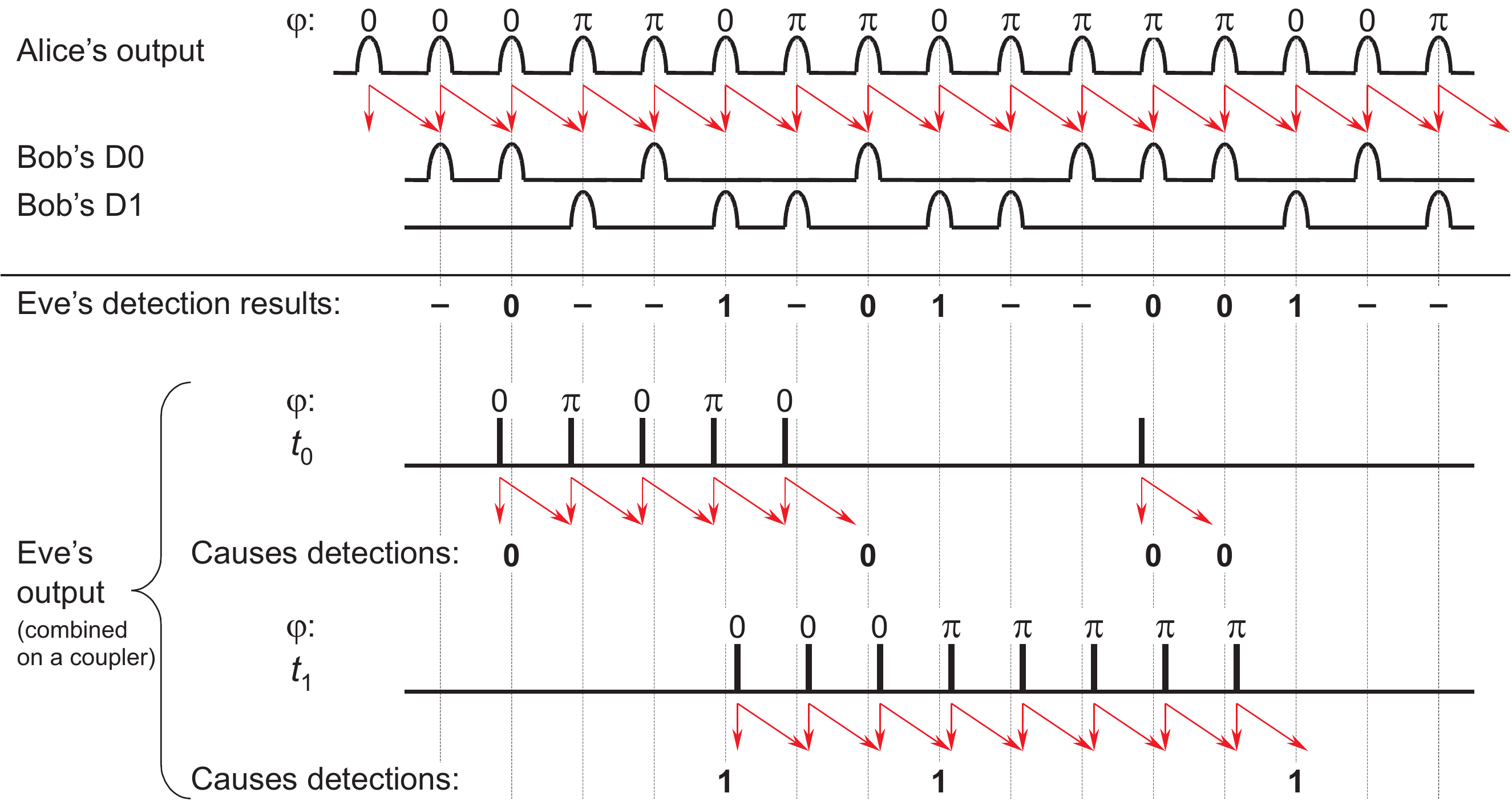}
\caption{\label{fig:dpsk-faked-states}Time diagram of a QKD system utilizing the DPSK ptotocol, and faked states for it. The three uppermost waveforms represent the intensity of light during normal system operation; the phase $\varphi$ of each Alice's pulse is noted. The rest of the diagram shows examples of possible faked states. For the compactness of illustration, Alice's average photon number per pulse is increased greatly, which gives Eve more frequent detections than would be possible in a real system. The arrows indicate how every pulse coming to Bob is split into the two arms of his interferometer.}
\end{figure*}

Faked states for this QKD system are constructed similarly to the previous one (Fig.~\ref{fig:dpsk-faked-states}).
In the case of the DPSK, Eve can run two generators of faked states in parallel, so that states with the values of the control parameter $t_0$ and $t_1$ may overlap. When Eve has had identical detection results in two adjacent bit slots, she can use a single-pulse faked state. In all other cases she generates longer faked states that encompass two or more detection results with the same bit value. In these faked states, Bob's other detector is blocked by the choice of the control parameter, and unwanted bit slots are blocked by destructive interference. In the limit, Eve may just generate two continuous trains of pulses with the control parameters $t_0$ and $t_1$, and modulate the phase of pulses in each train to produce the detections she wants at Bob.

In the system in Ref.~\onlinecite{takesue_2005}, Bob actually uses non-gated detectors, registers timing of all counts, and then selects timing ranges in software (this procedure is roughly equivalent to detector gating). In this system, Bob could easily implement monitoring of count statistics in the time domain, thus preventing Eve from using timing as a control parameter. However, we remind the reader that control parameters other than time can be used by~Eve. In this particular system, up-conversion detectors in Bob's setup employ narrow spectral filtering \cite{langrock_2005}. Eve could try to control wavelength of incoming pulses in addition to or instead of their timing.

The part of Eve's setup that generates faked states for both systems considered in this section may be similar to Alice's setup in Fig.~\ref{fig:DPSK-scheme}. In the case of the DPSK, two such setups could possibly be used, with their outputs combined on an optical coupler.

In the first of the two systems considered in this section, the system with the phase-time encoding, Bob would normally observe some coincidence counts at his detectors. To keep his coincidence rates the same as before the attack, Eve could occasionally simulate a coincidence count. She can do this by sending to Bob a faked state or several faked states that simultaneously address different detectors and/or bit slots. She could also control the photon number statistics of her faked states and employ the photon number measurement as described in the end of Sec.~\ref{sec:SARG04}.

Although we do not calculate it here, the faked states presented in this section would obviously work in the case of the partial efficiency mismatch, causing the more QBER the smaller the mismatch becomes. We note that schemes utilizing the DPSK protocol with limited-length states \cite{inoue_2002,buttler_2002} can also be attacked using the methods considered in this section.

\section{\label{sec:Ekert}Ekert protocol}

\begin{figure}
\includegraphics{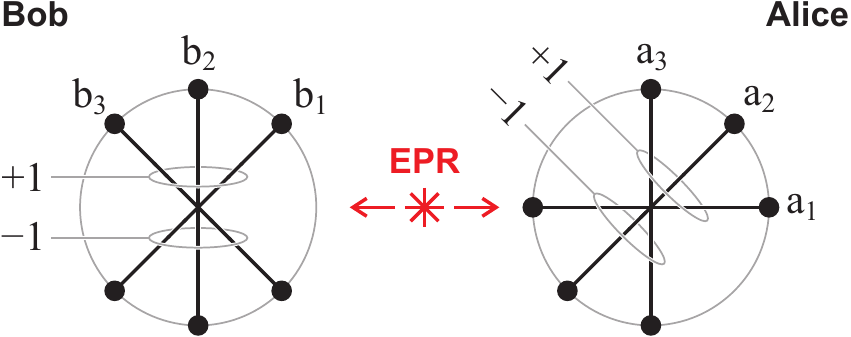}
\caption{\label{fig:Ekert-states}Possible measurements by Alice and Bob in the Ekert protocol. The circles represent the equator of the Poincare sphere. Measurement bases are denoted by letters with indices; each measurement can yield $+1$ or $-1$ result as labeled on the diagram. EPR, source of entangled photon pairs.}
\end{figure}

The Ekert protocol \cite{ekert} uses an external source of entangled pairs of photons in a singlet state, from which one photon is routed to Alice and the other to Bob. Alice and Bob perform measurements on their photons in one of the possible bases (Fig.~\ref{fig:Ekert-states}),
choosing between the bases randomly and independently of one another for each pair of incoming photons. After a series of measurements has taken place, the choices of bases are publicly announced. For those pairs where Alice and Bob both have registered a count in their detectors, quantum mechanics guarantees certain degree of correlation between the measurement results, depending on the combination of the bases chosen. The quantity
\begin{eqnarray}
\label{eq:ekert_E}
&E(\text{a}_i,\text{b}_j)=P_{++}(\text{a}_i,\text{b}_j)+P_{--}(\text{a}_i,\text{b}_j)\ \nonumber\\
&\ \ \ \ \ \ \ \ \ \ \ \ \ -P_{+-}(\text{a}_i,\text{b}_j)-P_{-+}(\text{a}_i,\text{b}_j)
\end{eqnarray}
is the correlation coefficient of the measurements performed by Alice in the $\text{a}_i$ basis and by Bob in the $\text{b}_j$ basis. Here $P_{\pm\pm}(\text{a}_i,\text{b}_j)$ denotes the probability that the result $\pm1$ has been obtained in the $\text{a}_i$ basis and $\pm1$ in the $\text{b}_j$ basis. For two identical pairs of bases ($\text{a}_2,\text{b}_1$ and $\text{a}_3,\text{b}_2$) the measurement results are totally anticorrelated:
\begin{equation}
\label{eq:ekert_anticorr}
E(\text{a}_2,\text{b}_1)=E(\text{a}_3,\text{b}_2)=-1.
\end{equation}
These measurement results are used in the protocol to form a secret key. Four other basis combinations are used to check for possible eavesdropping via computing the Clauser-Horne-Shimony-Holt quantity
\begin{equation}
\label{eq:ekert_S}
S=E(\text{a}_1,\text{b}_1)-E(\text{a}_1,\text{b}_3)+E(\text{a}_3,\text{b}_1)+E(\text{a}_3,\text{b}_3),
\end{equation}
which in the absence of eavesdropping should be equal to $-2\sqrt{2}$.

If the pairs of detectors on both Alice's and Bob's sides have a total efficiency mismatch, Eve can successfully mount a faked states attack that provides $S=-2\sqrt{2}$. She substitutes the source of entangled photons with one that generates, with certain probabilities, pairs of faked states listed below. We have assumed that, at Alice and at Bob, one detector is used to get the $+1$ measurement result in all three bases, and the other to get the $-1$ result. We have also assumed that Alice and Bob normalize detection probabilities separately for each combination of $\text{a}_i$ and $\text{b}_j$ before computing $E(\text{a}_i,\text{b}_j)$ correlation coefficients. Let's consider, step by step, how Eve can construct the faked states under these assumptions.

The simplest set of faked states necessary for the attack to work consists of two pairs of states; however, to make the $+1$ and $-1$ measurements on each side equiprobable, we'll be considering symmetric combinations consisting of two pairs of faked states each. The first combination named $\bm{\alpha}$ will be detected with a uniform probability and always produce total anticorrelation regardless of Alice's and Bob's choice of basis. It can, for example, consist of a pair of states conjugate to every other state used in the protocol and sent to Alice and Bob with opposite values of the control parameter $t_{+1}$ and $t_{-1}$, which blind the $-1$ and $+1$ detectors. When linear polarizations are used in the protocol, Eve can randomly send to Alice and Bob either a pair of circular polarizations $[(\text{circular})_{t_{+1}},(\text{circular})_{t_{-1}}]$ or a pair $[(\text{circular})_{t_{-1}},(\text{circular})_{t_{+1}}]$. If Eve only generated the combination $\bm{\alpha}$ and nothing else, it would result in
\begin{equation}
\label{eq:ekert_S_alpha}
S=-1-(-1)-1-1=-2.
\end{equation}
To reach the desired value of $S=-2\sqrt{2}$, we'll now target the second term in the equation for $S$. We devise a combination named $\bm{\beta}$ that only contributes to the $E(\text{a}_1,\text{b}_3)$ correlation coefficient but not to the other three correlation coefficients in the equation for $S$. In this combination, Eve sends either a pair $[(|{-}\text{a}_3\rangle)_{t_{+1}},(|{-}\text{b}_1\rangle)_{t_{+1}}]$ or a pair $[(|\text{a}_3\rangle)_{t_{-1}},(|\text{b}_1\rangle)_{t_{-1}}]$. It produces total correlation for the pair of bases $\text{a}_1,\text{b}_3$, as well as for three other pairs of bases ($\text{a}_1,\text{b}_2$; $\text{a}_2,\text{b}_2$; $\text{a}_2,\text{b}_3$) which are not used in the protocol. In the remaining five possible pairs of bases, the combination $\bm{\beta}$ causes no coincident detections. If the combinations $\bm{\alpha}$ and $\bm{\beta}$ are generated by Eve with probabilities $P_{\bm{\alpha}}=0.586$, $P_{\bm{\beta}}=0.414$, it results in
\begin{equation}
\label{eq:ekert_S_alpha_beta}
S=-1-(-0.172)-1-1=-2\sqrt{2}.\footnote{We make two remarks. Firstly, under the assumptions made, Eve could reach ``unphysical'' values of $S$ beyond $-2\sqrt{2}$ and almost up to $-4$ by increasing the weight $P_{\bm{\beta}}$ in her statistical mix. Secondly, now that we know how the states in the combination $\bm{\beta}$ look like, we can simplify Eve's apparatus by forming the combination $\bm{\alpha}$ of the same states. In the combination $\bm{\alpha}$, Eve can replace a circular polarization with a statistical mixture of two states from a single basis used in the protocol. In particular, she can send either a pair $[(|\text{a}_3\rangle \text{ or } |{-}\text{a}_3\rangle)_{t_{+1}},(|\text{b}_1\rangle \text{ or } |{-}\text{b}_1\rangle)_{t_{-1}}]$ or a pair $[(|\text{a}_3\rangle \text{ or } |{-}\text{a}_3\rangle)_{t_{-1}},(|\text{b}_1\rangle \text{ or } |{-}\text{b}_1\rangle)_{t_{+1}}]$.}
\end{equation}
Although we have reached the desired value of the quantity $S$, the terms in the equation for $S$ have unequal absolute values, which can be noticed by Alice and Bob. The absolute values of the terms can be made equal, just as they are in the absence of the attack, if we add a third combination. The third combination named $\bm{\gamma}$ contributes to all four correlation coefficients in the equation for $S$. In this combination, Eve sends either a pair $[(|{-}\text{a}_2\rangle)_{t_{+1}},(|{-}\text{b}_2\rangle)_{t_{+1}}]$ or a pair $[(|\text{a}_2\rangle)_{t_{-1}},(|\text{b}_2\rangle)_{t_{-1}}]$. It produces total correlation for the four pairs of bases used in computing $S$. It is easy to check that when the combinations are generated by Eve with probabilities $P_{\bm{\alpha}}=0.116$, $P_{\bm{\beta}}=0.653$, $P_{\bm{\gamma}}=0.231$, it results in
\begin{equation}
\label{eq:ekert_S_alpha_beta_gamma}
S=-0.707-(+0.707)-0.707-0.707=-2\sqrt{2}.
\end{equation}
Note that of the three combinations $\bm{\alpha}$, $\bm{\beta}$, $\bm{\gamma}$, only $\bm{\alpha}$ causes coincident detections in the pairs of bases $\text{a}_2,\text{b}_1$ and $\text{a}_3,\text{b}_2$ used to form the secret key. Detection results in these two pairs of bases are thus totally anticorrelated and the key is error-free. 

Although our attack reproduces the expected value of $S$, it has side effects. Detection probabilities for different combinations of bases become substantially unequal, and the three unused correlation coefficients are not reproduced properly. Thus, the attack relies on the absence of additional consistency checks on the data by the legitimate users. We have not been able to come up with a
set of faked states that does not produce any side effects. Also, the attack relies on the source of entangled photons being {\em outside} of Alice and Bob. If the source is placed inside one of their setups and only one of the two photons is accessible to Eve, it seems to us that with protocols that use more than two bases (the Ekert protocol and the six-state protocol \cite{bruss_sixstates_orig,bechmann-pasquinucci_1999,bennett_1984_ibm}), a zero-QBER attack using the approach described in this section cannot be constructed. However, the six-state protocol implemented on a setup that uses an external source of entangled photons could be successfully attacked using a faked pair source similar to the one described in this section.

\section{\label{sec:countermeasures}Countermeasures}

The partial detector efficiency mismatch is a flaw that is in principle unavoidable. Even if special care is taken to make detectors identical and eliminate possible control parameters, finite manufacturing precision will always leave possibility for Eve to control detector efficiencies to some extent. We therefore believe that the best approach to close this loophole is the following. Throughout the design, manufacture and quality assurance of the detectors and QKD system, the worst-case efficiency mismatch should be specified. It is possible that special measures would have to be taken to reduce the guaranteed value of mismatch (for example, in gated detectors it can be introduction of a random jitter into the detector gating signal). Then, the worst-case value of efficiency mismatch should be accounted in the general security proof for the protocol used, and the amount of privacy amplification if necessary be increased to guarantee security of the key. For the BB84 protocol, some quantitative estimates for the extra privacy amplification exist (see Sec.~\ref{sec:BB84}); for other protocols, they are not yet known.

Monitoring the bit rates and coincidence statistics for different bit-basis combinations is useful as a general precaution. It is good because it can make Eve's life more difficult, as well as monitor the health of the hardware better. However, as we have discussed, Eve might have ways to maintain the bit rate and coincidence statistics unchanged, so this measure does not guarantee security.

A countermeasure has recently been proposed in which Bob randomly switches assignment of his detectors to 0 and 1 bit values by applying an additional $\pi$ shift at his phase modulator \cite{QuantInfComp-7-p73}. For example, in the BB84 protocol Bob would randomly apply one of the four phase shifts ($-\frac{3\pi}{4}$, $-\frac{\pi}{4}$, $\frac{\pi}{4}$, $\frac{3\pi}{4}$) at his modulator to choose a combination of detection basis and detector assignment, instead of two phase shifts ($-\frac{\pi}{4}$, $\frac{\pi}{4}$) to choose the detection basis. This countermeasure would prevent the straightforward faked states attack, because Eve would not know how to construct the faked state without knowing the assignment of detectors in advance. However, Eve could run a time-shift type attack \cite{QuantInfComp-7-p73,arXiv-0704-3253v1-quant-ph} in combination with the large-pulse attack against Bob that reads his phase modulator settings \cite{large-pulse-attack}. In the time-shift attack, Eve only needs to know the assignment of detectors {\em after} she has manipulated quantum states. It is in practice difficult to protect Bob's modulator from external interrogation, because any additional protective optical components at Bob's input would introduce unwanted attenuation to quantum states. Thus, this countermeasure does not seem to be sufficient.

The cases of efficiency mismatch considered in this paper are necessarily idealized. There are many modifications to the setups that would break the described attacks, e.g., using a slightly wider gate for one detector than for the other, or having four detectors in the setup instead of two. However, such modifications do not eliminate efficiency mismatch per se, and the problem that Eve might still exploit it (even if it is a one-sided mismatch) using a more sophisticated attack remains.
%\\% the final touch that fixes TeX layout bug

\section{\label{sec:conclusion}Conclusion}

We have shown that detector efficiency mismatch can be exploited to attack the SARG04 and Ekert protocols, as well as schemes that use the phase-time encoding and the DPSK protocol. The faked states attacks considered here might not be the optimal ones; however, they certainly set upper bounds on the secret information. We emphasize the necessity of characterizing the detector setup thoroughly and establishing security proofs with partial detector efficiency mismatch integrated into the equipment model.

\bibliography{paper}% Produces the bibliography via BibTeX.

\end{document}